\documentclass[twoside,twocolumn,9pt]{article}
\usepackage{extsizes}
\usepackage[super,sort&compress,comma]{natbib} 
\usepackage[version=3]{mhchem}
\usepackage[left=1.5cm, right=1.5cm, top=1.785cm, bottom=2.0cm]{geometry}
\usepackage{balance}
\usepackage{times,mathptmx}
\usepackage{sectsty}
\usepackage{graphicx} 
\usepackage{lastpage}
\usepackage[format=plain,justification=justified,singlelinecheck=false,font={stretch=1.125,small,sf},labelfont=bf,labelsep=space]{caption}
\usepackage{float}
\usepackage{fancyhdr}
\usepackage{fnpos}
\usepackage{multicol}
\usepackage[english]{babel}
\usepackage{soul}
\addto{\captionsenglish}{%
  
}
\usepackage{array}
\usepackage{droidsans}
\usepackage{charter}
\usepackage[T1]{fontenc}
\usepackage[usenames,dvipsnames]{xcolor}
\usepackage{setspace}
\usepackage[compact]{titlesec}
\usepackage{hyperref}

\usepackage{epstopdf}

\definecolor{cream}{RGB}{222,217,201}

\begin{document}
\pagestyle{fancy}
\thispagestyle{plain}
\fancypagestyle{plain}{

\renewcommand{\headrulewidth}{0pt}
}

\makeFNbottom
\makeatletter
\renewcommand\LARGE{\@setfontsize\LARGE{15pt}{17}}
\renewcommand\Large{\@setfontsize\Large{12pt}{14}}
\renewcommand\large{\@setfontsize\large{10pt}{12}}
\renewcommand\footnotesize{\@setfontsize\footnotesize{7pt}{10}}
\renewcommand\scriptsize{\@setfontsize\scriptsize{7pt}{7}}
\makeatother

\renewcommand{\thefootnote}{\fnsymbol{footnote}}
\renewcommand\footnoterule{\vspace*{1pt}%
\color{cream}\hrule width 3.5in height 0.4pt \color{black} \vspace*{5pt}} 
\setcounter{secnumdepth}{5}

\makeatletter 
\renewcommand\@biblabel[1]{#1}            
\renewcommand\@makefntext[1]%
{\noindent\makebox[0pt][r]{\@thefnmark\,}#1}
\makeatother 
\renewcommand{\figurename}{\small{Fig.}~}
\sectionfont{\sffamily\Large}
\subsectionfont{\normalsize}
\subsubsectionfont{\bf}
\setstretch{1.125} 
\setlength{\skip\footins}{0.8cm}
\setlength{\footnotesep}{0.25cm}
\setlength{\jot}{10pt}
\titlespacing*{\section}{0pt}{4pt}{4pt}
\titlespacing*{\subsection}{0pt}{15pt}{1pt}

\fancyfoot{}
\fancyfoot[RO]{\footnotesize{\sffamily{1--\pageref{LastPage} ~\textbar  \hspace{2pt}\thepage}}}
\fancyfoot[LE]{\footnotesize{\sffamily{\thepage~\textbar\hspace{3.45cm} 1--\pageref{LastPage}}}}
\fancyhead{}
\renewcommand{\headrulewidth}{0pt} 
\renewcommand{\footrulewidth}{0pt}
\setlength{\arrayrulewidth}{1pt}
\setlength{\columnsep}{6.5mm}
\setlength\bibsep{1pt}

\makeatletter 
\newlength{\figrulesep} 
\setlength{\figrulesep}{0.5\textfloatsep} 

\newcommand{\topfigrule}{\vspace*{-1pt}%
\noindent{\color{cream}\rule[-\figrulesep]{\columnwidth}{1.5pt}} }

\newcommand{\botfigrule}{\vspace*{-2pt}%
\noindent{\color{cream}\rule[\figrulesep]{\columnwidth}{1.5pt}} }

\newcommand{\dblfigrule}{\vspace*{-1pt}%
\noindent{\color{cream}\rule[-\figrulesep]{\textwidth}{1.5pt}} }

\makeatother

\twocolumn[
  \begin{@twocolumnfalse}
\vspace{3cm}
\sffamily
\begin{tabular}{m{4.5cm} p{13.5cm} }

& \noindent\LARGE{\textbf{Anomalous thermal fluctuation distribution sustains proto-metabolic cycles and biomolecule synthesis$^\dag$}} \\
 & \vspace{0.3cm} \\

 & \noindent\large{Rowena Ball,$^{\ast}$\textit{$^{a}$} John Brindley\textit{$^{b}$} } \\

\end{tabular}

 \end{@twocolumnfalse} \vspace{0.6cm}

  ]

\renewcommand*\rmdefault{bch}\normalfont\upshape
\rmfamily
\section*{}
\vspace{-1cm}


\footnotetext{\textit{$^{a}$~Mathematical Sciences Institute and Research School of Chemistry, Australian National University, Canberra, ACT 2602 Australia; E-mail: Rowena.Ball@anu.edu.au}}
\footnotetext{\textit{$^{b}$~School of Mathematics, University of Leeds, Leeds LS2 9JT UK. 
}}

\footnotetext{\dag~Electronic Supplementary Information (ESI) available:  See DOI: 00.0000/00000000.}




\sffamily{\textbf{An environment far from equilibrium is thought to be a necessary condition for the origin and persistence of life. In this context we report open-flow simulations of a non-enzymic proto-metabolic system, in which hydrogen peroxide acts both as oxidant and driver of thermochemical cycling. We find that a Gaussian perturbed input produces a non-Boltzmann output fluctuation distribution around the mean oscillation maximum. Our main result is that net biosynthesis can occur under fluctuating cyclical but not steady drive. Consequently we may revise the necessary condition to ``dynamically far from equilibrium''.}}\\


\rmfamily 


It is generally assumed that primeval chemical evolution and the origin of life must have taken place in a strongly nonequilibrium milieu \cite{Goppel:2016} which also, and inevitably, was subject to random perturbations, or some degree of noisiness and variability \cite{Nisbet:2001,Mast:2010,Brogioli:2011,Filisetti:2011,Szostak:2011,Wu:2012,Branscomb:2017,Guttenberg:2018,Matsubara:2018}. Regular heating-cooling cycles are also understood to have been necessary to drive polymerisation, replication and proto-metabolism in pre-cellular proto-living systems \cite{Lahav:1978,Keil:2016,Kitadai:2018}. 
Two relevant questions arising naturally are then: (I) Could some form  of thermal fluctuation distribution, produced by random perturbations of the environmental  temperature,  drive proto-metabolic cycles and biosynthetic processes advantageously over  hydrolysis and degradation reactions? (II) In particular, how do Gaussian perturbations of the inputs affect the outputs from a thermally periodic drive? 

In previous articles we reported simulation studies of replication of a small RNA  \cite{Ball:2015} and  a prebiotic-relevant dimerization \cite{Ball:2017,Ball:2019} driven by the thiosulfate-hydrogen peroxide (THP) thermochemical-pH oscillator (see Table S1 in the ESI$^\dag$). For the dimerization we found that hetereogeneity and stochastic inputs  led to a favorable distribution of fluctuations of the thermal oscillation amplitude; in other words, for that case the answer to question (I) above is ``yes''. Here we report an important generalisation of those results to more complex chemical systems. We also, in response to question (II), explore the more fundamental physics of the connection between input and output temperature fluctuations for open complex reaction systems.

We find empirically that the frequency distribution of output temperature fluctuations due to Gaussian thermal fluctuations of the input is non-Gaussian, and in nonequilibrium steady state is skewed and weighted so that a complex chemical system under such governance inevitably is degraded, or cannot develop.  But this life-denying scenario changes dramatically under dynamical conditions of self-sustained thermochemical cycling. By isolating the frequency distribution of fluctuations around the mean oscillation amplitude  we show how the fundamental physics that governs thermochemical relaxation oscillations, associated with the specific heat of the liquid medium,  also permits the sustainment and growth of complex chemical systems, and that it  may manifest in transiently non-Boltzmann populations of molecular and intermolecular quantized modes. 

The influence of spontaneous concentration  fluctuations on simple replicating networks in the isothermal case was studied by Brogioli \cite{Brogioli:2011}. 
He proposed that such networks must drift along a manifold of marginally stable stationary states in order to evolve spontaneously. Our results relax that stringent requirement in the more realistic, nonisothermal scenario where thermal cycling drives a reaction network. 

\begin{figure} 
\centering
\includegraphics[scale=1]{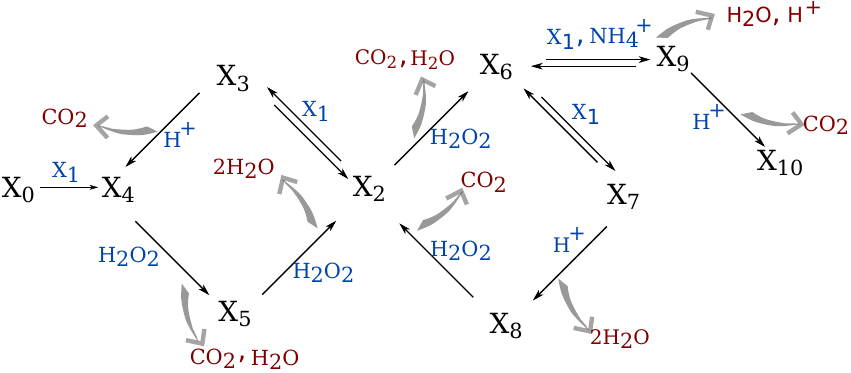} 
\caption{\label{figure1}  Schematic of the proto-metabolic system from   \cite{Springsteen:2018} as used in this work. {\rm X}$_1$ is glyoxalate,  {\rm X}$_2$ is oxaloacetate,  {\rm X}$_{10}$ is aspartate, the other species are identified in the ESI$^\dag$. 
} 
\end{figure}

We used the TCA (tricarboxylic acid) metabolic cycle analogue that was proposed and studied experimentally by Springsteen \textit{et al.} \cite{Springsteen:2018}. The experimenters used hydrogen peroxide as the oxidant and thermostatted temperature changes, the objective being to show how a simple metabolic cycle could operate in the absence of redox enzymes in an origin-of-life scenario. Fig. \ref{figure1} shows the reaction system of two oxidative cycles fed by glyoxalate, X$_1$, and linked via oxaloacetate, X$_2$, that ultimately drive production of aspartate, X$_{10}$. We used the  THP oscillator to drive this system in open flow, thus providing both the hydrogen peroxide and temperature cycling smoothly and self-consistently. 

 The simulated system consists of the proto-metabolic cycle reactions of Fig. \ref{figure1} plus those of the THP oscillator (see ESI$^\dag$), housed in a nonequilibrium flow-through cell coupled to a thermal bath, with parameters set in the range over which thermochemical oscillations are spontaneously generated and sustained.  
It is modelled by the following dynamical mass balances, 
\begin{equation}
V\frac{\text{d}c_x}{\text{d}t}=V\sum_p \sigma r_{i} (T) + F(c_{x,\text{f}} - c_x), \label{e1}
\end{equation}
where $c_x$ is  concentration of reactant or intermediate $x$, the summation  is over the $p$ reaction rates $r_{i, (i=1\ldots p)}$ that involve  $x$,  $\sigma=1$ for intermediates  produced and $\sigma=-1$ for reactants and intermediates  consumed, $V$ is the cell volume and  $F$ is the volumetric flow rate, 
and the enthalpy balance 
\begin{equation}
V\bar{C}\frac{\text{d}T}{\text{d}t} = V\sum_n (-\Delta H_i) r_i (T) + (F\bar{C})(\tilde{T}_\text{a} - T) + L(\tilde{T}_\text{a} - T), \label{e2}
\end{equation}
 where  the summation is over the $n$ reaction enthalpy  $(-\Delta H_i)$ contributions, $\bar{C}$ is the average volumetric specific heat, and  $L$ is the thermal conductance of the cell walls. 
Equations \eqref{e1} and \eqref{e2} are thermally coupled through the Arrhenius temperature dependences of the coefficients $k_i$ of the reaction rates $r_i$,  where
$
k_i(T) = z_i\exp\left(\frac{-E_i}{RT}\right)
$ and $z_i$ and $E_i$ are the frequency factors and activation energies, $R$ is the gas constant. 
The stochastic ambient (or thermal bath) temperature is $\tilde{T}_\text{a}= \bar{T}_\text{a} + \delta T_\text{a}$, where $\bar{T}_\text{a}$ is the mean and $\delta T_\text{a}$ is the normally distributed random fluctuation with variance $s=\omega  \bar{T}_\text{a}$. This mimics the necessarily messy environment that must have hosted pre- and proto-cellular chemical evolution. 
Equations \eqref{e1} and \eqref{e2} were integrated to obtain the output temperature and concentrations.  

\begin{figure*}
\centering
\includegraphics[scale=1]{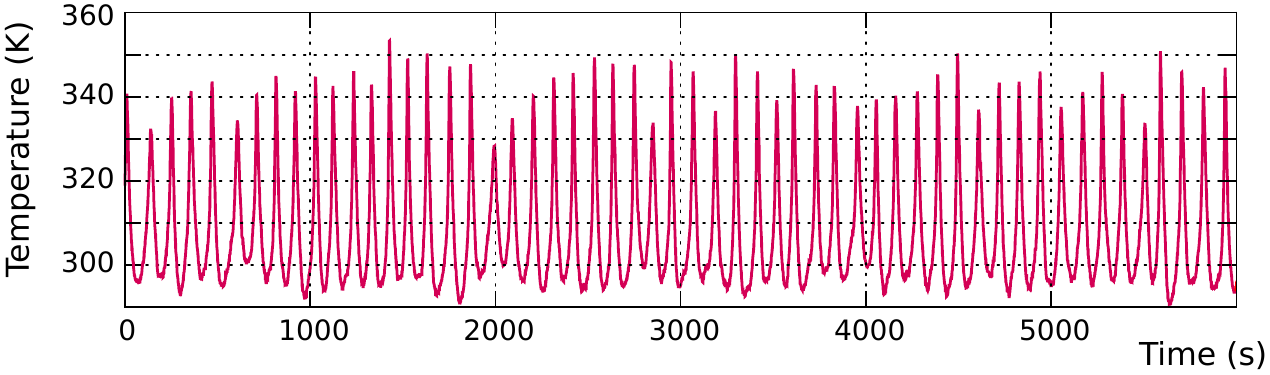} 
\caption{\label{figure2} A time window of the computed cell temperature from Eqs \eqref{e1}--\eqref{e2}.} 
\end{figure*}

A typical time series within the oscillatory  {regime} is shown in Fig. \ref{figure2}. The values of the input parameters  {$\bar{T}_\text{a}$, $L$, $\bar{C}$ and the average output temperature $T$ and aspartate concentration ${\text{X}}_{10}$} are given as Case I in Table \ref{table1}. The fluctuations of the oscillation maxima in Fig. \ref{figure2} are notable --- these are purely a dynamical phenomenon.  A great many smaller fluctuations on the troughs, shoulders, upsweeps and downsweeps also are evident on close inspection --- these are a steady state phenomenon. 

\begin{figure*}[h] 
\centering
\includegraphics[scale=.9]{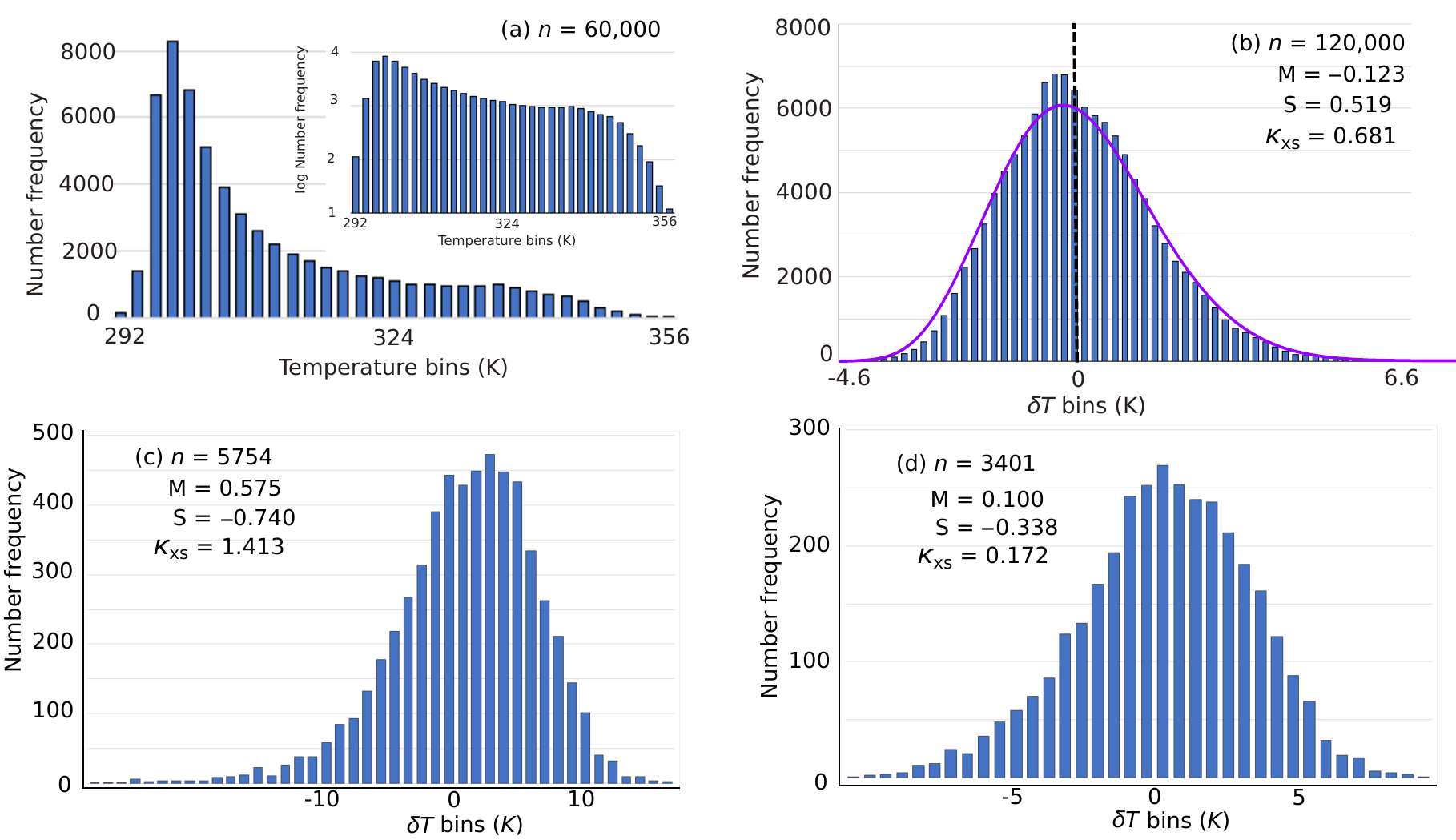} 
\caption{\label{figure3} Frequency histograms, where \textit{n} number of data points, M median, S skewness, $\kappa_{\text{xs}}$ excess kurtosis. (a)  The fluctuating temperature from the time series, extended, of Fig. \ref{figure2}. (b)--(d) Output temperature fluctuations: (b)  in steady state with Bezier-approximated envelope, (c) and (d) in dynamical state --- fluctuations of the thermal oscillation mean maximum for the full system (c) and  for the reactions of the THP oscillator alone (d).
} 
\end{figure*}

We summarize  the  fluctuations in the frequency histogram  shown in Fig. \ref{figure3} (a). Although resulting from normally distributed perturbations of the ambient temperature, the distribution is  manifestly  bimodal with a maximum at around 298\,K and a maximum of lower number frequency at around 336\,K. 

\begin{table}
\caption{\label{table1}  Four cases treated in the text
}
\centerline{\small
\begin{tabular}{p{0.02\textwidth}p{0.03\textwidth} p{0.07\textwidth}p{0.07\textwidth}p{0.03\textwidth}p{0.08\textwidth}}
\hline
Case  & $\bar{T}_\text{a}$\,(K) & $L$\,(mW\,K$^{-1}$) & $\bar{C}$\,(J(K\,L)$^{-1}$) &$T$\,(K) & ${\text{X}}_{10}$\,{(M)}\\\hline
I & 282 & 0.002& 3400 & $^*$308 & $^*7.62 \times 10^{-8}$\\
II & 282 & 0.0025 & 3400& 296 & $3.34 \time 10^{-10}$\\
III& 290 & 0.0035 & 3400 & 308 & $4.11\time 10^{-9}$\\
IV &282 & 0.002 & 4400 & 296 & $3.51 \times 10^{-10}$\\
 \hline
\end{tabular}
}
\small $^*$time-average value
\end{table}

We can observe the high number frequency, low temperature, distribution without interference by damping the  THP oscillations; this is achieved simply by increasing the thermal conductance $L$ until (mathematically) a Hopf bifurcation point is crossed, and taking the histogram of another time series.  (Note that this method is physically preferable  to taking an algebraic steady state.) 
This is Case II in Table \ref{table1}.  The histogram is shown in Fig.~\ref{figure3}~(b) where we have used the perturbation $\delta T$ around the mean of zero. Interestingly, it has the form of a gamma probability distribution, which characteristically has positive skewness and positive excess kurtosis.  It shows that  large high-temperature fluctuations occur with low frequency and  is weighted towards lower temperatures.  A complex molecular system under its governance must degrade into simpler elements over time,  because in typical biochemical condensations, such as those of Fig. \ref{figure1}, the non-enzymic degradation or hydrolysis generally has a lower activation energy than the synthesis or condensation.  
 Such a system cannot  sustain complex biochemical cycles and chemical evolution and inevitably must revert to a flowing soup of simple species, unable to grow or evolve. 

In Case III, Table \ref{table1}, $\bar{T}_\text{a}$ and $L$ are adjusted within the fluctuating steady state {regime} to obtain an average output temperature  equal to that of Case I,  but the output concentration of aspartate is lower  by a factor of 20.  

It is known that in a closed, non-reactive, near-equilibrium thermodynamic system at steady state the fluctuations induced  by Gaussian thermal fluctuations through the boundary are non-Gaussian, and are described by a gamma probability density function~\cite{Bertola:2015}. 
By the numerical distribution in Fig.~\ref{figure3}~(b) and its calculated statistical parameters  this remains true for an open, reactive, far-from-equilibrium system at steady state. 
Clearly, then, the ``strongly out of equilibrium'' condition, though necessary, is not sufficient for life, and we may infer that the dynamical distribution of fluctuations --- the smaller hump at higher temperature in Fig. \ref{figure3} (a) --- also is needed to sustain proto-metabolic cycles and biosynthetic processes. What does it look like?

We have picked out the thermal oscillation maxima from the time series, extended, in Fig. \ref{figure2} and show the resulting histogram of perturbations around the mean in Fig. \ref{figure3} (c). For comparison in Fig. \ref{figure3} (d) we show the similar histogram for a run of the THP oscillator reactions alone.  
We elicit several important points.
\begin{enumerate}
\item The calculated statistical parameters are given but, really, visual  inspection of the distribution in Fig.~\ref{figure3}~(c) tells the story clearly: Over time, high temperature, high activation energy synthetic reactions must be favoured over the reverse, low activation energy, hydrolysis and degradation reactions, thus complex processes leading to life are enabled. 
\item A right-weighted, left-skewed distribution  is evident in Fig.~\ref{figure3}~(d) as well as in Fig.~\ref{figure3}~(c),  thus we infer that the THP oscillator provides a suitable drive for proto-metabolic processes and chemical to biological evolution.  
\item The additional nonlinearity and complexity provided by the metabolic reactions right-weights and left-skews the distribution in Fig. \ref{figure3} (c) further than that in (d). Given that there is an optimum fluctuation distribution (which we do not know precisely) that balances metabolic and synthetic reactions against destructive effects of high temperature maxima that may occur too often, this suggests that the chemical system itself selects this distribution to optimize performance, in this case, the output of aspartate.  The long-term effect  is to produce more structure, and more complex structure, and allow greater persistence of structure \cite{Bejan:2011}.  

\end{enumerate}

The above discussion having cast some light on questions (I) and (II), we are led immediately to question (III): What is the mechanism behind the dynamical distributions  in Fig.~\ref{figure3} (c) {and (d)}, an extraordinary contrast to that  in Fig.~\ref{figure3} (b), even though both are produced by strongly nonequilibrium systems?  

For a fluctuating steady state the fluctuation distribution  reflects the physics behind the equilibrium Boltzmann distribution of internal quantized modes  and to this extent it is not surprising that we see the gamma distribution in Fig. \ref{figure3} (b). But this physics does not govern dynamical fluctuations, in this case, of the oscillation amplitude.  

The dynamical  behaviour of a thermoreactive system is governed by the specific heat of the medium, $\bar{C}$, which multiplies the dynamical term on the left-hand side of Eq. \ref{e2}. In liquid media the major contributor, $\sim$90\%, to the specific heat is the potential energy of the intermolecular vibrational and intramolecular rotational modes \cite{Ball:2011}.  

Thermochemical oscillations are true relaxation oscillations  and their physics in terms of the specific heat of the medium was elucidated by two-timing analysis\cite{Ball:2011}. Their characteristic form is non-sinusoidal and non-symmetric. 
Typically, in ``slow time'' reactant is being consumed but the temperature is barely rising, because the reaction enthalpy is stored in the internal quantized potentials of the medium. A temperature spike occurs when these modes become saturated.   At the temperature peak the ``fast'', dissipative time scale becomes dominant and the temperature relaxes,  linearly and independently of the specific heat, because the reactant (hydrogen peroxide in this case)  locally is depleted. We may make an educated guess  that the internal mode populations  become transiently non-Boltzmann just before the explosive heat release and temperature spike, and this population inversion may be reflected in the shape of the distributions  in Fig.~\ref{figure3} (c)  {and (d)}. 

How reasonable is this ansatz, granted that the specific heat is modelled as a bulk, constant quantity in Eqs \eqref{e1} and \eqref{e2}? 
Several cases of reactive chemical systems have been reported where the reaction enthalpy deposited into rotational and vibrational modes produces non-Boltzmann populations at moderate temperatures \cite{Kim:1999,Tomellini:2003}. More generally 
it has been shown that nonequilibrium systems with sufficiently complex stochastic dynamics may exhibit non-Boltzmann statistics \cite{Beck:2003}. Such systems can exhibit a superposition of two (or more) different statistics, or superstatistics: the foundational one given by ordinary Boltzmann factors, and another given by the fluctuations  at a different local thermal state. The superstatistics of our system  is exemplified in Fig. \ref{figure3} (a), which is a superposition of the steady-state distribution shown in (b), governed by the Boltzmann population distribution of internal quantized modes, and  the dynamical distribution in (c), governed by the specific heat of the medium and the extent to which non-Boltzmann populations of those modes can be supported. 

{An alternative method of damping the dynamics  of Eqs }\eqref{e1}--\eqref{e2}{  is to vary the  specific heat $\bar{C}$ until a Hopf bifurcation condition fails }\cite{Ball:2011}{. Physically this means that the fluctuation distribution of Fig. }\ref{figure3}{ (c) is suppressed and that of Fig.~}\ref{figure3}{~(b) remains,  so that  the high activation energy metabolic and synthetic reactions (such as aldol condensations) cannot be sustained over the reverse reactions of lower activation energy.} The resulting relatively poor production of aspartate in a medium of (unrealistically) high specific heat is given as Case IV in Table \ref{table1}. 

From the results presented in Fig. \ref{figure3} and these discussions we are led to the following conclusions: 
\begin{enumerate}
\item {The far-from-equilibrium requirement is a necessary but not sufficient condition for stochastic media to support nonenzymic proto-metabolism, biosynthesis and chemical evolution. Additionally, such media are required to sustain non-steady, or dynamical activity that can produce the right-weighted, left-skewed (or ``pro-life'') spectrum of thermal fluctuations exemplified in Fig. }\ref{figure3}{ (c), which on average supports high activation energy biosynthetic reactions over low activation energy degradation reactions. Nonequilibrium stochastic media  at steady state cannot support such processes because the  fluctuation spectrum is left-weighted and right-skewed (or ``anti-life'') as in Fig. }\ref{figure3}{  (b), so that over time low activation energy degradation reactions prevail.}
\item These results provide constraints on the types of media in which chemical evolution is likely to occur. Postulates that extraterrestrial life may be found in seas of liquid hydrocarbons, such as on Titan, are limited by the low specific heat of such media. At the other end, the high specific heat of fresh water may preclude it as a medium for chemical evolution. The dissolved salts in seawater and the presence of hydrogen peroxide may bring the specific heat into the conducive  ``goldilocks'' range.  
\item It is likely that the anomalous fluctuation frequency distributions in Fig. \ref{figure3} (c) {and (d)} are reflected from transient non-Boltzmann populations of quantum states of the liquid medium. We suggest that experiments designed to observe  low frequency intermolecular vibrations  in the  
far infrared in a dynamical THP medium may prove rewarding. 
\end{enumerate} 

 We make the general observation arising from and illustrated by this work that the study of thermal fluctuation distributions is a powerful but under-utilized tool for furthering understanding of the {behaviour} of complex nonequilibrium chemical systems. 

\section*{\normalsize Conflicts of interest}
There are no conflicts to declare.
\section*{\normalsize Acknowledgements}
This research was partially supported by Australian Research Council Future Fellowship  FT0991007 for R. B. 

\providecommand*{\mcitethebibliography}{\thebibliography}
\csname @ifundefined\endcsname{endmcitethebibliography}
{\let\endmcitethebibliography\endthebibliography}{}

   \end{document}


\title{Supplementary Information for\\ `Anomalous thermal fluctuation distribution sustains proto-metabolic cycles and biomolecule synthesis'}

\author{Rowena Ball }
\affiliation{Mathematical Sciences Institute and Research School of Chemistry, Australian National University, Canberra, ACT 2602 Australia }

\author{John Brindley} 
\affiliation{School of Mathematics, University of Leeds, Leeds LS2 9JT UK }

\begin{abstract}\bigskip
Included in this Supplementary Information are all data necessary to reproduce (and build on) the results presented in the associated article. Choices for numerical integration and data post-processing software are left to individual researchers.
\end{abstract}

   \maketitle

\bigskip\bigskip
\noindent \raggedright{Table S1. Reactions (1)--(4) are the oxidations and acid-base equilibria of the minimal THP oscillator. Reactions (5)--(15) are those of the linked metabolic cycles and aspartate synthesis in Fig. (1) of the paper. X$_0$ is pyruvate, X$_1$ is glyoxalate, X$_2$ is oxaloacetate, X$_3$ is oxalomalate, X$_4$ is 4-hydroxy-2-ketoglutarate, X$_5$ is malate, X$_6$ is malonate, X$_7$ is 3-carboxy-malate, X$_8$ is 3-carboxy-oxaloacetate, X$_9$ is pyruvate, X$_{10}$ is aspartate.} 
\begin{align}
\text{H}_2\text{O}_2 + \text{HSO}_3^- & \longrightarrow \text{SO}_4^{2-} + \text{H}_2\text{O} + \text{H}^+\\
\text{S}_2\text{O}_3^{2-} + 2\text{H}_2\text{O}_2 & \longrightarrow \frac{1}{2}\text{S}_3\text{O}_6^{2-}  +  \frac{1}{2} \text{SO}_4^{2-} +2 \text{H}_2\text{O}\\
 \text{HSO}_3^- &\rightleftarrows  \text{H}^+ + \text{SO}_3^{2-}\\
  \text{H}_2\text{O} &\rightleftarrows  \text{H}^+ + \text{OH}^-\\
   \text{X}_0^- +  \text{X}_1^- &\rightleftarrows   \text{X}_4^{2-} \\
  \text{X}_1^- + \text{X}_2^{2-} &\rightleftarrows  \text{X}_3^{3-} \\
   \text{X}_3^{3-} + \text{H}^+  & \longrightarrow  \text{X}_4^{2-} + \text{CO}_2 \\
  \text{X}_4^{2-} +  \text{H}_2\text{O}_2  & \longrightarrow  \text{X}_5^{2-} + \text{CO}_2 +  \text{H}_2\text{O} \\
 \text{X}_5^{2-} +    \text{H}_2\text{O}_2  & \longrightarrow  \text{X}_2^{2-} + 2 \text{H}_2\text{O} \\
  \text{X}_2^{2-}+ \text{H}_2\text{O}_2 & \longrightarrow   \text{X}_6^{2-} + \text{CO}_2 +  \text{H}_2\text{O} \\
   \text{X}_6^{2-}  +  \text{X}_1^- &\rightleftarrows   \text{X}_7^{3-} \\
  \text{X}_7^{3-} +   \text{H}_2\text{O}_2 & \longrightarrow  \text{X}_8^{3-} + 2 \text{H}_2\text{O}\\
   \text{X}_8^{3-} +  \text{H}^+ & \longrightarrow  \text{X}_2^{2-}  + \text{CO}_2\\
    \text{X}_6^{2-}  +   \text{X}_1^-  + \text{NH}_4^+ & \rightleftarrows   \text{X}_9^{3-} + \text{H}^+  + \text{H}_2\text{O}\\
     \text{X}_9^{3-} + \text{H}^+ & \longrightarrow \text{X}_{10} +  \text{CO}_2
\end{align}

\newpage

\raggedright{Table S2. Thermokinetic parameters for reactions (1)--(15) and reaction enthalpies for reactions (1)--(4). The specific enthalpies of reactions (5)--(15) are negligible.  \\Due to the scarcity of published thermokinetic data for the organic reactions (5)--15),  in the simulations we used rate parameters based on values for reactions involving similar functional groups and substituents or deduced on the basis of textbook chemical kinetics principles \cite{Laidler:1987,Connors:1990}. 
We found that the system behaviour is robust to minor changes in the kinetic parameters for  reactions (5)--(15). \\$^*$Units as appropriate. }\\[2mm]

\centerline{
\begin{tabular}{p{0.1\textwidth}p{0.16\textwidth} p{0.3\textwidth}p{0.26\textwidth}}
\hline
     & $E$ (kJ/mol) & $z^*$ & $\Delta H$\,(kJ/mol)\\\hline
(1) & 35.0 & $2.08 \times 10^6$ & $-380.0$ \\
 (2) & 68.1 & $1.63\times 10^{10} $ & $-572.2$\\
(3) & 75.1, 25.8 & $3.95 \times 10^{16}$, $1.61\times 10^{15}$& 49.3, $-49.3$\\
(4) & 55.8, 0 & $4.55\times 10^6$, $1.32\times 10 ^9$ & 55.8, $-55.8$\\
(5) & 58.0, 59.0 & $7.7\times 10^{10}$, $2.7\times 10^8$ \\
(6) & 58.0, 59.0 & $7.7\times 10^{10}$, $2.7\times 10^8$ \\
(7) &58.0 & $7.7\times 10^{10}$ \\
(8) & 68.0 & $7.7\times 10^{10}$  \\
(9) & 58.0 & $7.7\times 10^{10}$ \\
(10) & 68.0 & $7.7\times 10^{10}$  \\ 
(11) & 58.0, 59.0 & $1.2\times 10^{10}$, $1.4\times 10^{8}$ \\
(12) & 68.0 & $7.7\times 10^{10}$ \\
(13) & 63.0 & $7.7\times 10^{10}$  \\
(14) & 68.0, 70.0 & $7.7\times 10^{10}$, $2.7\times 10^8$  \\
(15) & 63.0 & $7.7\times 10^{10}$ \\
 \hline
\end{tabular}
} 

\bigskip\bigskip\bigskip
Table S3. Values of the feed concentrations $c_{x,\text{f}}$,  flow rate $F$ and cell volume  $V$ used in Eqs (1) and (2) of the article. \\[2mm]

\centerline{
\begin{tabular}[b]{>{\raggedright}>{\raggedleft}p{0.2\textwidth}p{0.17\textwidth}p{0.17\textwidth}p{0.17\textwidth}}
\hline
	$F$\,(L\,s$^{-1}$) & & $4.8\times 10^{-7}$ \\
	$V$\,(L)	 && $1\times 10^{-5}$ \\
	$c_{x,\text{f}}$\,(M)			 &H$_2$O$_2$ & 1.35 & \\
				& $\text{S}_2\text{O}_3^{2-}$ & 0.9 \\
				& H$^+$ & $2\times 10^{-7}$ \\
				& $ \text{SO}_3^{2-}$ & $1.6\times 10^{-3}$ \\
				& $\text{X}_1^-$ & 0.27 \\
				& $\text{X}_0^-$ & 0.14\\
				& $ \text{NH}_4^+$ & 0.1\\	
\hline
 \end{tabular}
}

\bigskip

    \bibliographystyle{rsc}
   \bibliography{mc}